\documentclass[]{elsart}
\usepackage{epsfig}
\usepackage{amssymb}
  \def \cima   {ClO$^-_2$-I$^-$-MA }
  \def \cdima  {ClO$_2$-I$_2$-MA }

  \def \amat {\mbox{\boldmath $A$}}
  \def \lmat {\mbox{\boldmath ${\mathcal L}$}}
  \def \mmat {\mbox{\boldmath $M$}}
  \def \smat {\mbox{\boldmath $S$}}
  \def \imat {\mbox{\boldmath $I$}}
  \def \Cvect {\mbox{\boldmath ${\mathcal C}$}}
  \def \Fvect {\mbox{\boldmath ${\mathcal F}$}}
  \def \fvect {\mbox{\boldmath $f$}}
  \def \Rvect {\mbox{\boldmath $R$}}
  \def \zvect {\mbox{\boldmath $\zeta$}}
  \def \Dmat  {\mbox{\boldmath $D$}}
  \def \calDmat  {\mbox{\boldmath ${\mathcal D}$}}
  \def \df    {\mbox{\boldmath $df$}}
  \def \dg    {\mbox{\boldmath $dg$}}
  \def \Uvect {\mbox{\boldmath $U$}}	
  \def \Psivect   {\mbox{\boldmath $\Psi$}}	
\begin{document}
\begin{frontmatter}
\title{Exploring Defective Eigenvalue Problems with the Method of Lifting}
\author[caltech]{S. Setayeshgar\corauthref{cor}\thanksref{now}},
\corauth[cor]{Corresponding author.}
\thanks[now]{Present address: Department of Physics, Princeton University, Princeton, NJ 08544;  
Tel: (609) 258-4320, FAX: (609) 258-1124, E-mail: simas@math.princeton.edu}
\author[caltech]{H. B. Keller},
\address[caltech]{Department of Applied Mathematics, California Institute of Technology,
Pasadena, California 91125} 
\author[lanl]{J. E. Pearson}
\address[lanl]{XCM, MS F645, Los Alamos National Laboratory, Los Alamos, New 
Mexico 87545}

\begin{abstract}
Consider an $N \times N$ matrix \amat \, for which zero is a defective
eigenvalue.  In this case, the
algebraic multiplicity of the zero eigenvalue is greater
than the geometric multiplicity.  We show how 
an \textsl{inflated} $(N+1) \times (N+1)$
matrix \lmat \, can be constructed as a rank one perturbation to
\amat \,, such that \lmat \, is singular but no
longer defective, and the nullvectors of \lmat \, 
can be easily related to the nullvectors of \amat.  
The motivation for this construction 
comes from linear stability analysis of
an experimental reaction-diffusion system which
exhibits the Turing instability.
The utility of this scheme is accurate
numerical computation of nullvector(s) corresponding to a
defective zero eigenvalue.  We show that numerical computations
on \lmat \, yield more accurate eigenvectors than direct
computation on \amat.
\end{abstract}

\end{frontmatter}


\section{Introduction}
\label{sec:intro}

In this paper, we present a method for improving the accuracy of computed
eigenvalues and eigenvectors of defective matrices 
\cite{ref:texts}.  This algorithm, which
we refer to as \textsl{lifting}, involves construction of a nondefective
matrix in a higher dimensional space.  Standard linear algebra
procedures are shown to yield more accurate numerical results for the lifted matrix.
Our scheme formally resembles the construction of an augmented system in pseudo-arclength
continuation \cite{ref:HBK1,ref:HBK2,ref:HBK3,ref:HBKtext}.  Its formulation was motivated 
by the structure of the linear stability
analysis of a particular chemical reaction-diffusion system, known as the 
chlorine-dioxide--iodine--malonic-acid system 
\cite{ref:P_92,ref:PB_92,ref:SS_MCC_98,ref:SS_MCC_99} (henceforth referred to
as CDIMA), which exhibits the Turing instability.

First, we present the \textsl{Lifting Theorem}.  
We demonstrate its numerical implementation on ``small''
and ``large'' defective eigenproblems.  We describe the connection with the 
CDIMA linear stability problem, and discuss the general applicability of 
this scheme.

\section{Lifting}

An eigenvalue is defective if its algebraic multiplicity is greater than
its geometric multiplicity.  (The \textsl{geometric multiplicity} of an
eigenvalue equals the dimension of the space spanned by the eigenvectors
corresponding to that eigenvalue.)  Consider an eigenvalue with geometric
multiplicity equal to one: It is nondefective if the corresponding
left and right eigenvectors are nonorthogonal.  For a defective eigenvalue, lifting
corresponds to a geometric procedure of casting the corresponding  
eigenvectors of the matrix and the adjoint matrix 
into a higher dimensional space in which these vectors are no longer orthogonal.
In this section, we show how to construct such a well-conditioned matrix in 
the higher dimensional space from a defective matrix.    

For simplicity, we define the notation:
\begin{eqnarray*}
\mbox{algebraic multiplicity of eigenvalue $\mu$ of matrix \mmat \,} 
                      & \equiv & AM(\mmat; \mu) \\
\mbox{geometric multiplicity of eigenvalue $\mu$ of matrix \mmat \,}
                      & \equiv & GM(\mmat; \mu).
\end{eqnarray*}
Although in the following we consider zero eigenvalues and
corresponding nullvectors, the discussion holds generally for nonzero
eigenvalues, with $\amat \,=\mmat \, - \mu \imat_N \,$, where $\imat_N$
is the appropriate $N \times N$ identity matrix.

\vspace{.2in}
\noindent
{\bf Theorem:} Consider an $N \times N$ matrix \amat, such that zero
is a defective eigenvalue of \amat \, with $AM(\amat; 0)=n$ and
$GM(\amat; 0)=1$, $n>1$. The right and left unit
nullvectors of \amat \, are given by {\boldmath $\phi$} and 
{\boldmath $\psi^\top$}, respectively.  An $(N+1) \times (N+1)$ 
matrix \lmat \, constructed according to
\begin{equation}
\lmat \, = \left(\begin{array}{cc}
                      \amat & {\bf 0} \\
		      {\bf 0}^\top & 0
		   \end{array} \right) + \mbox{\boldmath $v w^\top$}. 
\end{equation}
where
\begin{equation}
\mbox{\boldmath $v$} = \left( \begin{array}{c}
		      {\bf v} \\
		      \eta 
		 \end{array} \right), \qquad
\mbox{\boldmath $w$} = \left( \begin{array}{c}
		      {\bf w} \\
		      \omega 
		 \end{array} \right),
\end{equation}
will have a simple zero eigenvalue if
\begin{eqnarray}
\mbox{(i)} \;\; {\bf w}^\top \mbox {\boldmath $\phi$} & \neq &0,
\label{eq:cond1} \\
\mbox{(ii)} \;\;  \mbox {\boldmath $\psi$}^\top {\bf v} & \neq & 0, 
\label{eq:cond2} \\
\mbox{(iii)}  \;\;\;\;\;  \eta \omega & \neq & 0.
\label{eq:cond3}
\end{eqnarray}
${\bf v}$ and ${\bf w}$ are $N$-dimensional vectors, and $\eta$ and $\omega$
are scalars.

\vspace{.2in}
\noindent
{\bf Proof:} First, we show there exists a nontrivial 
vector {\boldmath $\Phi$}, given by
\begin{equation}
\mbox {\boldmath $\Phi$} = \left( \begin{array}{c}
		      {\bf x} \\
		      \xi
		 \end{array} \right),
\end{equation}
such that
\begin{equation}
	\lmat \, \mbox {\boldmath $\Phi$} = {\bf 0}.
\end{equation}
For {\boldmath $\Phi$} to be a right nullvector of \lmat, the
$N$ components of $ \left(\lmat \, \mbox {\boldmath $\Phi$} \right)$ in the
\textsl{original} space, as well as the $(N+1)^{th}$ component
in the \textsl{inflated} space must be zero:
\begin{eqnarray}
\amat \, {\bf x} + \alpha {\bf v} &=&0 \label{eq:origcomps}\\
\alpha \eta &=&0 \label{eq:inflatedcomp}, 
\end{eqnarray}
where 
\begin{equation}
\label{eq:alphadef}
\alpha= {\bf w}^\top {\bf x} + \omega \xi.
\end{equation}
Since $\eta \neq 0$, for Eq.\ \ref{eq:inflatedcomp} to be satisfied,
we must have $\alpha = 0$.
Then, Eq.\ \ref{eq:origcomps} implies
${\bf x} = r \mbox{\boldmath $\phi$}$, where $r$ is a constant,
and $\xi$ must be given by
\begin{equation}
\xi = - r \, \frac{{\bf w}^\top \mbox {\boldmath $\phi$}}{\omega}.
\label{eq:xi}
\end{equation}
Hence, the pointwise projection of the right nullvector of \lmat \,
onto the original $N$-dimensional space is the right nullvector
of \amat, up to a multiplicative constant.\footnote{If {\boldmath $\Phi$}
is normalized to give the unit right nullvector, then 
$r^2 = \left(1+{\bf w}^\top \mbox {\boldmath $\phi$}/\omega\right)^{-1}$.}
Additionally, its $(N+1)^{th}$ component, $\xi$, is
nonzero.

Likewise, we can show that \lmat \, has a nontrivial left nullvector
{\boldmath $\Psi^\top$} given by
\begin{equation}
\mbox {\boldmath $\Psi$} = \left( \begin{array}{c}
		      {\bf y} \\
		      \zeta
		 \end{array} \right),
\end{equation}
where
\begin{equation}
{\bf y} = s \mbox{ \boldmath$\psi$}, \qquad
\zeta = - s \, \frac{\mbox {\boldmath $\psi^\top$} {\bf v}}{\eta}. 
\end{equation}
Again, $s$ is a constant, and we have $\zeta \neq 0$.

Next, we must show that zero is a \textsl{simple} eigenvalue of \lmat.
This will be done in two steps: 
\begin{eqnarray}
&{\rm (i)}& \quad GM(\lmat;0)=1, \nonumber \\
&{\rm (ii)}& \quad \mbox{\boldmath $\Psi^\top \Phi 
                      = \psi^\top \phi$} + \zeta \xi \neq 0.
\label{eq:tildeinnerproduct}
\end{eqnarray} 
First, we prove $GM(\lmat;0)=1$ by contradiction.  Assume $GM(\lmat;0)>1$.
Then, there exists $\mbox{\boldmath $\Phi^\prime \neq \Phi$}$ such that
\begin{equation}
\lmat \, \mbox {\boldmath $\Phi^\prime$} = {\bf 0}, \qquad
\mbox {\boldmath $\Phi^\prime$} = \left( \begin{array}{c}
		      {\bf x^\prime} \\
		      \xi^\prime
		 \end{array} \right).
\end{equation}
Eqs.\ \ref{eq:origcomps} and \ref{eq:inflatedcomp} imply 
$\amat \, {\bf x^\prime} = {\bf 0}$.
We distinguish two cases:
\begin{itemize}
\item[{a.}] If ${\bf x^\prime} \neq r \mbox{\boldmath $\phi$}$, 
then ${\bf x^\prime}$
is a right nullvector of \amat, different from $r \mbox {\boldmath $\phi$}$.
This violates the assumption $GM(\amat;0)$=1.
\item[{b.}]If ${\bf x^\prime} = r \mbox{\boldmath $\phi$}$, 
we must have $\xi^\prime \neq \xi$ 
for $\mbox{\boldmath $\Phi^\prime \neq \Phi$}$ .  
However, from Eqs.\ \ref{eq:inflatedcomp} and \ref{eq:alphadef},
$\xi^\prime = - r {\bf w}^\top \mbox {\boldmath $\phi$}/\omega = \xi$.
Therefore, $\mbox{\boldmath $\Phi^\prime$}$ is not different
from $\mbox{\boldmath $\Phi$}$.
\end{itemize}
Hence, we have shown that $GM(\lmat;0)=GM(\amat;0)=1$.  (This
proof is easily generalized to $GM(\lmat;0)=GM(\amat;0)=\nu$.)
To show (ii), we note that zero being a defective eigenvalue of 
\amat \, with $GM(\amat;0)=1$ implies
$\mbox{\boldmath $\psi^\top \phi$}=0$.  However, the inner product 
given by Eq.\ \ref{eq:tildeinnerproduct} will still be nonzero, 
since $\zeta \xi \neq 0$. This completes the proof that
zero is a simple eigenvalue of \lmat.

It is important to consider the generic
case in numerical applications,  
where \amat \, is not exactly defective
but almost-defective, $\mbox{\boldmath $\psi^\top \phi$} \approx 0$. 
For zero to be a nondefective eigenvalue of \lmat,
the inner product given by Eq.\ \ref{eq:tildeinnerproduct} 
must be nonzero:
\begin{equation}
\mbox{\boldmath $\psi^\top \phi$} + 
\left( \frac{\mbox {\boldmath $\psi^\top$} {\bf v}}{\eta} \right)
\left( \frac{{\bf w}^\top \mbox {\boldmath $\phi$}}{\omega} \right) \neq 0.
\label{eq:additionalcond}
\end{equation}
For example, constructing \lmat \, with \mbox{\boldmath $v$}
and \mbox{\boldmath $w$} such that ${\bf v}=
\mbox{\boldmath $\phi$}$, $\eta=-1$, and 
$\omega={\bf w}^\top \mbox {\boldmath $\phi$}$, would 
violate the condition above. 
Therefore, when \amat \, is almost-defective, Eq.\ 
\ref{eq:additionalcond} must be additionally satisfied to guarantee
that \lmat \, will not be defective.  Below, we give a general prescription
for choosing the lifting vectors, which unlike the pathological special
case just presented, should avoid construction of a
defective lifted matrix.

Consider choosing (for simplicity) $\eta=\omega=1$, and 
$\left( {\bf v}, {\bf w} \right) = \left( \beta {\bf v_r}, \gamma {\bf w_r} \right)$: 
\begin{equation}
{\rm v_r}_i={\tt rand}(i)/N_{\rm v}, \qquad {\rm w_r}_j={\tt rand}(j)/N_{\rm w},
\end{equation}
where ${\tt rand}(i)$ are random numbers on 
$\left[ -1 , 1 \right ]$, $N_{\rm v}$ and $N_{\rm w}$ are
normalizing factors leading to unit random vectors, and $\gamma$ and $\beta$
are constants.  This construction should guarantee that 
when \amat \, is singular and exactly defective 
(with $AM(\amat; 0)=n$, $GM(\amat; 0)=1$), or close to defective,
then zero will be a simple eigenvalue of \lmat.
We will refer to the constants $\gamma$ and $\beta$ as \textsl
{lifting parameters}, since the $(N+1)^{th}$ components of the
left/right nullvectors of $\lmat$ are proportional to these
constants, respectively, and measure the projections of these
vectors onto the new, inflated direction.

To extend the above theorem and proof to the case where \amat \, is
defective and $AM(\amat; 0) > GM(\amat; 0) > 1$,
we must show that \lmat \,  can be constructed such that zero is a 
nondefective, multiple eigenvalue 
with $AM(\lmat; 0)=GM(\lmat; 0)=GM(\amat; 0)$. With multiple
zero eigenvalues, \lmat \, will be defective if there exists
a vector ${\bf g}$ such that 
\begin{equation}
	\lmat {\bf g}=\mbox{\boldmath $\Phi_i$},
\label{eq:genvect}
\end{equation}
for some $\mbox{\boldmath $\Phi_i$} \in {\mathcal N}(\lmat)$, where
$i=1 \ldots \nu$, and ${\mathcal N}(\lmat)$ is the nullspace of \lmat.
(If such a vector ${\bf g}$ existed, then it would be a generalized
eigenvector of \lmat, associated with an $m \times m$ Jordan block,
where $m>1$.)  Taking the inner product of Eq.\ \ref{eq:genvect}
with $\mbox{\boldmath $\Psi_j^\top$}$, where
$\mbox{\boldmath $\Psi_j^\top$} \in {\mathcal N}(\lmat^\top)$, 
$j=1 \ldots \nu$, gives
\begin{equation}
\mbox{\boldmath $\Psi_j^\top$} \mbox{\boldmath $\Phi_i$} = 0.
\end{equation}
Therefore, for \lmat \, to be nondefective, 
it must be constructed such that no
vector in the right nullspace of \lmat \, is orthogonal
to its entire left nullspace. 


\section{Numerical Results}
\subsection{$2 \times 2$ Defective Eigenproblem}
Consider the following matrix, \mmat:
\begin{equation}
\mmat \, = \left(\begin{array}{cc}
                      \pi & 1 \\
		      -\pi^2/4 & \epsilon
		   \end{array} \right). 
\label{eq:mmat2}
\end{equation}
The eigenvalues of \mmat \, are given by
\begin{equation}
\mu_{\pm}=\left[\pi+\epsilon \pm \sqrt{ \left( \epsilon^2-2 \pi \epsilon \right)} \right]/2.
\end{equation}
The two components of an eigenvector {\boldmath $\phi^\top$}$=(\phi_1, \phi_2)$
satisfy
\begin{equation}
\phi_2/\phi_1 = \mu - \pi.
\end{equation}
where $\mu=\mu_\pm$.
For $\epsilon=0$, \mmat \, is defective.  We use the lifting
algorithm to explore the error in
computing the eigenvector associated with $\mu_+$ when \mmat \, is close
to being defective, $\epsilon \ll 1$.  
The lifted matrix \lmat \, is constructed with 
$\amat =\mmat  - \mu_+ \imat_2$ according to
\begin{equation}
\label{eq:constructionii}
\mbox{\boldmath $v$} = \left( \begin{array}{c}
		      \beta {\bf v_r} \\
                      1 
		 \end{array} \right), \qquad
\mbox{\boldmath $w$} = \left( \begin{array}{c}
		      \beta {\bf w_r} \\
		      1
		 \end{array} \right),
\end{equation}
where $({\bf v_r, w_r})$ are two-dimensional random unit
vectors, and we set $\gamma = \beta$ without loss of generality. 

The right unit nullvector \mbox{\boldmath $\Phi$} of \lmat \, is computed
using MATLAB's {\tt eig()} function \cite{ref:MATLAB,ref:Matlab_guide}.
We explore the lifting error, constructed according to
\begin{equation}
{\mathcal E}(\epsilon, \beta)=\left | \Phi_2/\Phi_1 - (\mu_+ - \pi) \right|.
\end{equation}
as a function of parameters $\epsilon$ and $\beta$.

\begin{figure}
   \centering \leavevmode
   \epsfxsize=5.5in
   \epsfbox{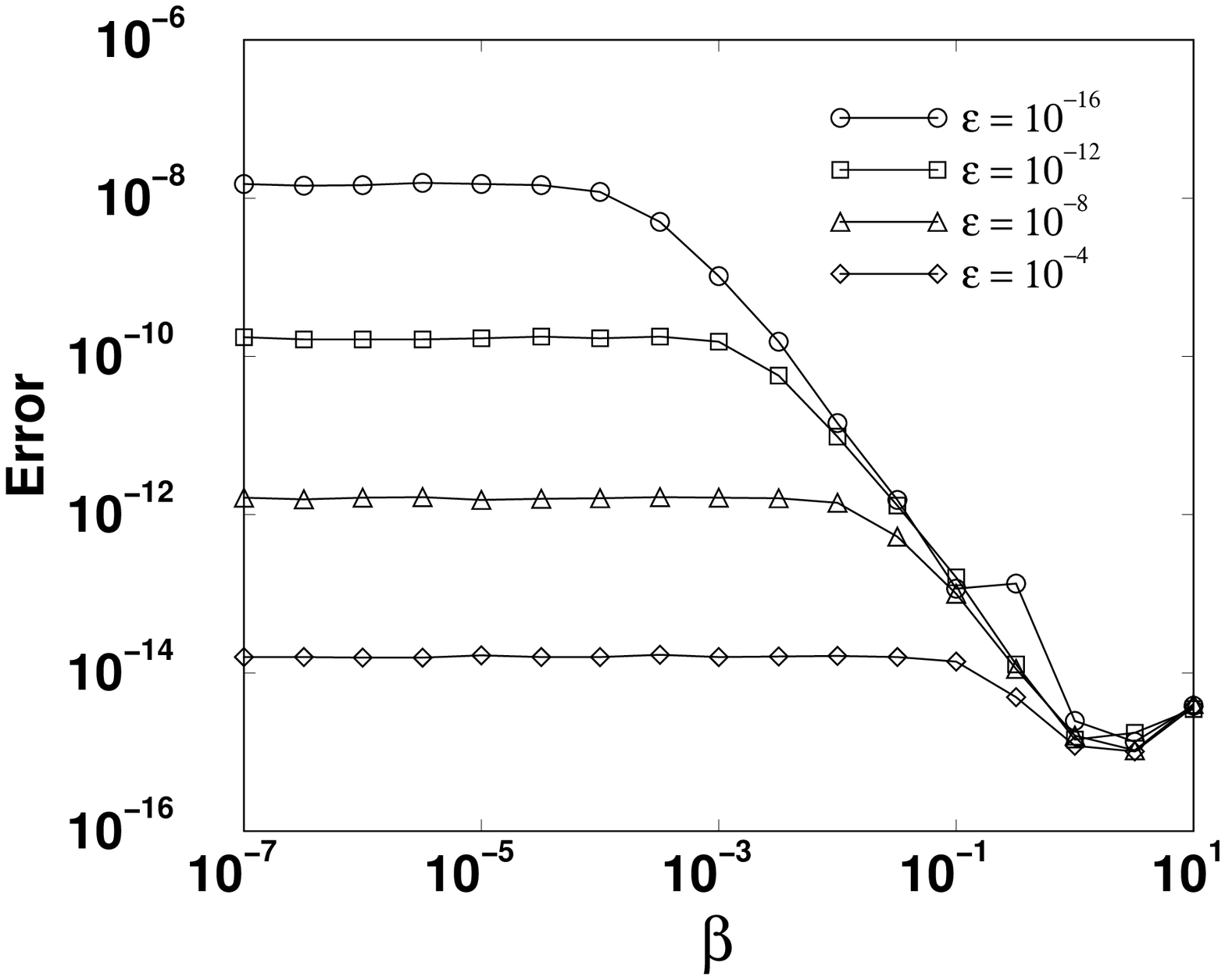}
   \vspace{0.2in}
   \caption[]{\small
The lifting error, ${\mathcal E}(\epsilon, \beta)$, is plotted as
a function of lifting parameter, $\beta$, for different values
of $\epsilon$.  
The lifting vectors are given by: $ \mbox{\boldmath$v$} = (\beta \, {\bf v_r}, 1)$
and $ \mbox{\boldmath$w$} = (\beta \, {\bf w_r}, 1)$, 
where ${\bf v_r}$ and ${\bf w_r}$ are two-dimensional unit random
vectors.  Each point corresponds to the mean error obtained using
$1000$ unit random lifting vectors, $({\bf v_r, w_r})$.  
}
   \label{fig:smallmat_eps_random}
\end{figure}

\begin{figure}
   \centering \leavevmode
   \epsfxsize=5.5in
   \epsfbox{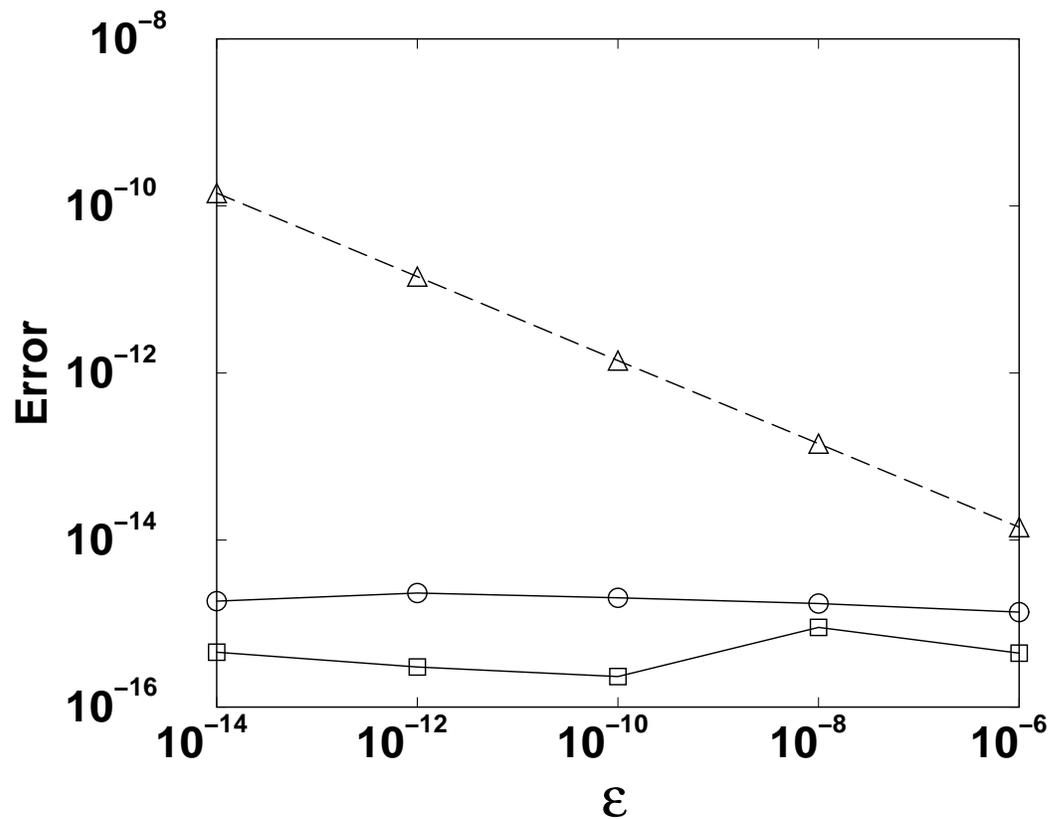}
   \vspace{0.2in}
   \caption[]{\small
Error as a function of $\epsilon$ with lifting parameter $\beta=1.0$: 
Open circles show the error using lifting
vectors given by $ \mbox{\boldmath$v$} = (\beta \, {\bf v_r}, 1)$
and $ \mbox{\boldmath$w$} = (\beta \, {\bf w_r}, 1)$, 
where ${\bf v_r}$ and ${\bf w_r}$ are two-dimensional unit random
vectors.  Each point corresponds to the mean error obtained using $1000$ unit
random lifting vectors, $({\bf v_r, w_r})$.
Open squares show the error using lifting vectors given by
$ \mbox{\boldmath$v$} = (\beta \, \mbox{\boldmath$\psi$}, 1)$
and $\mbox{\boldmath$w$} = (\beta \, \mbox{\boldmath$\phi$}, 1)$, 
where $\mbox{\boldmath$\psi$}$ and $\mbox{\boldmath$\phi$}$ are the left and right
unit nullvectors of $\amat$, respectively.  
The error without lifting is given by the long-dashed line (open triangles).  

}
   \label{fig:smallmat_eps_all}
\end{figure}

Figure\ \ref{fig:smallmat_eps_random}
shows ${\mathcal E}(\epsilon, \beta)$ as a function of the 
lifting parameter $\beta$ for different values of $\epsilon$.  
The basic trend is that of decreasing error as a function of increasing
$\beta$ for small to moderate values of $\beta$.
In this figure, each point corresponds to the mean
error computed using $1000$ unit random lifting vectors,
$({\bf v_r, w_r})$.  The root-mean-square (rms) of the distribution of errors
is of the order of the mean error for small values of lifting
parameter.  
For $\beta \,\mbox{\raisebox{-0.6ex}{$\stackrel{>}{\sim}$}} \, {\mathcal O}(1)$,
the rms is at most an order of magnitude greater
than the mean error for values of $\epsilon$ close to machine precision;  
for larger values of $\epsilon$, the rms
is equal to a few times the mean error.  We have verified that
the larger values of rms  are a result of a few outliers.
In Figure\ \ref{fig:smallmat_eps_all}, for lifting parameter $\beta =1.0$, 
we compare the mean error with lifting (open circles) and without lifting (open triangles)
as a function of $\epsilon$.  In this figure, we also show 
the error using lifting vectors given by 
({\bf v, w})$=(\mbox{\boldmath$\psi$}, \mbox{\boldmath$\phi$})$, 
such that the nonorthogonality conditions
Eqs.\ \ref{eq:cond1} and \ref{eq:cond2} are satisfied by definition (open squares).
We note that the lifting error is within an order of magnitude of
machine precision for all values of $\epsilon$.  

\begin{figure}
   \centering \leavevmode
   \epsfxsize=5.5in
   \epsfbox{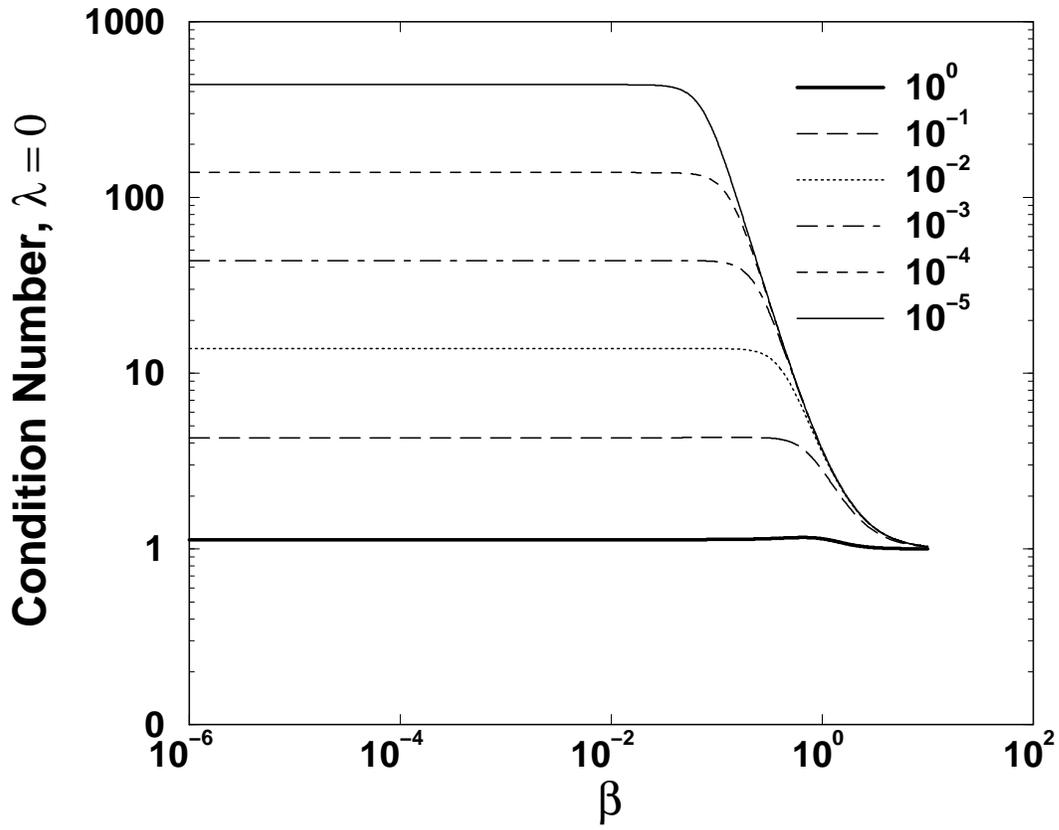}
   \vspace{0.2in}
   \caption[]{\small
The condition number for the singular value of the $3 \times 3$
lifted matrix, \lmat \,, is plotted as
a function of lifting parameter, $\beta$, for different values
of $\epsilon$.  A single, unit random lifting vector, $({\bf v_r, w_r})$,
was used to construct the lifted matrix.
}
   \label{fig:condnum_random}
\end{figure}

The condition number of the simple singular value of the lifted matrix is
given by the reciprocal of
\begin{equation}
	s(0) = \left| \mbox{\boldmath $\Psi$}^\dagger \mbox{\boldmath $\Phi$} \right|,
\end{equation}
where \mbox{\boldmath $\Psi$} and \mbox{\boldmath $\Phi$} are the left and
right unit nullvectors of \lmat.  In Figure\ \ref{fig:condnum_random}, we show the
condition number as a function of the lifting parameter $\beta$ for different
values of $\epsilon$, using a single pair of unit random lifting vectors:  
For $\epsilon \ll 1$,  the condition number clearly improves with
increasing lifting parameter.
 
\subsection{$N \times N$ Defective Eigenproblem}

For applications in which the matrix
is a discrete approximation to a continuous operator,
\amat \, will have dimension $N \gg 1$.  To
explore the method of lifting in computing the defective eigenvector 
of larger matrices, we constructed such an example from the $2 \times 2$
matrix above: with $\mmat$ 
in the top left block, along
with the $(N-2) \times (N-2)$ block tridiagonal matrix from Poisson's
equation in the bottom right, we applied a similarity transformation to the
resulting $N \times N$ matrix to obtain a general matrix with
no special properties (such as symmetry or bandedness).  The
eigenvalue, $\mu_+$, of this matrix is
exactly known, and the lifted matrix \lmat \, is constructed as 
in the $2 \times 2$ case.

\begin{figure}
   \centering \leavevmode
   \epsfxsize=5.5in
   \epsfbox{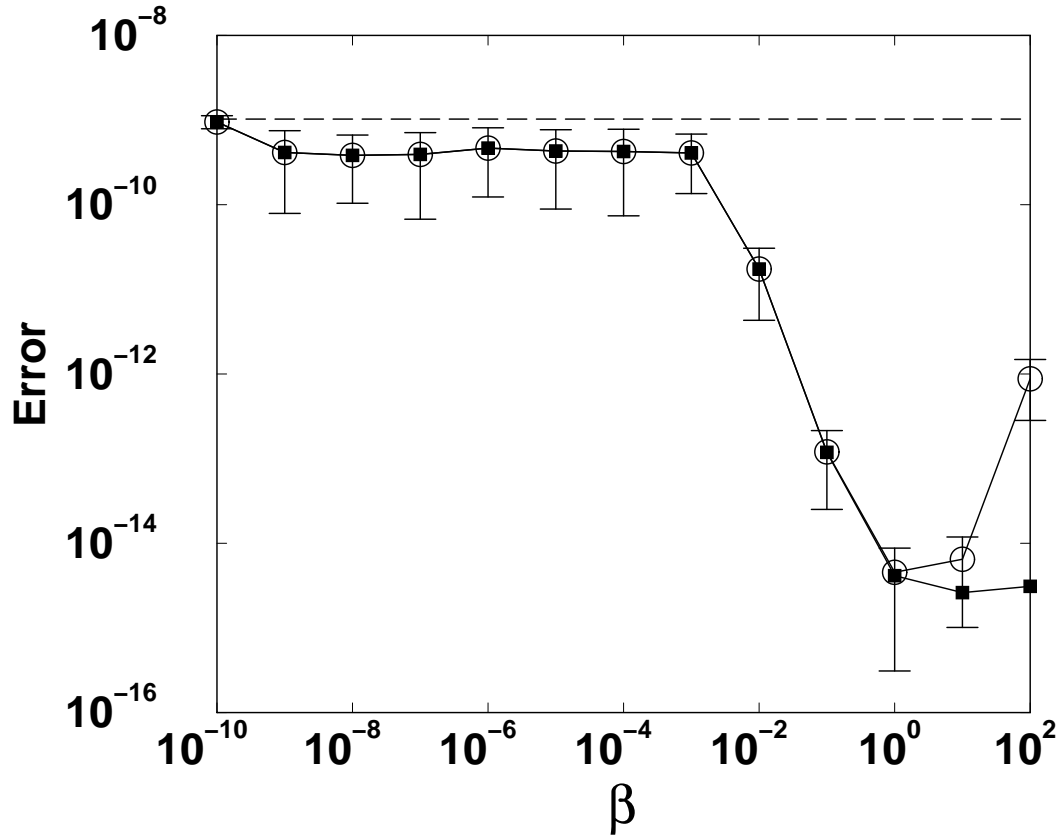}
   \vspace{0.2in}
   \caption[]{\small
The lifting error (open circles) and the magnitude of the ``zero''
eigenvalue, $\left| \lambda_0 \right|$,
of \lmat \, (filled squares) are plotted as a function of
lifting parameter, $\beta$, with $\epsilon=10^{-12}$ and $N=500$.
The error without lifting is shown for comparison (long-dashed line).  The lifting
vectors are given by: $\mbox{\boldmath$v$} = (\beta \, {\bf v_r}, 1)$
and $\mbox{\boldmath$w$} = (\beta \, {\bf w_r}, 1)$, where ${\bf v_r}$ and 
${\bf w_r}$ are $(N-1)$-dimensional unit random
vectors.  Each point corresponds to the average quantity 
obtained using $50$ unit random
lifting vectors, $({\bf v_r, w_r})$, and the error bars give the rms 
of the distribution of errors.
}
   \label{fig:largemat_random}
\end{figure}

Once the right nullvector of \lmat, given by $\mbox{\boldmath $\Phi^\top$}=
\left({\bf x}, \xi \right)$ is numerically computed, 
the lifting error is constructed as before:
\begin{equation}  
{\mathcal E}(\epsilon, \beta) = \left |\right(\smat {\bf x} \left)_2/\right(\smat {\bf x} \left)_1
                       -(\mu_+ - \pi)\right|,
\end{equation}
(where \smat \, is the similarity transformation).
In Figure\ \ref{fig:largemat_random}, we show the mean error and its rms
computed using $50$ unit random lifting vectors, $({\bf v_r, w_r})$. 
We find that for small values of $\beta$, the lifting error is
dominated by the error in computing the ``zero'' eigenvalue, $\lambda_0$, of \lmat.
For $\beta \,\mbox{\raisebox{-0.6ex}{$\stackrel{>}{\sim}$}} \, {\mathcal O}(1)$, 
the magnitude of this eigenvalue remains at machine
precision, while the lifting error is larger.
For values of $\beta$ such that the magnitude of the ``zero''
eigenvalue is at machine precision, the lifting error is dominated
by the error in computing the associated eigenvector.  
We note that there is an ``optimal'' value of $\beta$ giving
the smallest lifting error, which is of order unity.  For values
of $\beta$ much larger than unity, the error increases 
as a function of increasing $\beta$.


\section{CDIMA Eigenproblem}

The symmetry-breaking instability of a system
from a homogeneous steady state to a patterned state, predicted in
1952 by Alan Turing \cite{ref:Tur_52}, was observed for the first time in the
chlorite-iodide-malonic acid \cima (CIMA) reaction-diffusion system 
\cite{ref:Castets_90,ref:Swinney_91}.  
In practice, due to boundary chemical feeds, the steady
state is not homogeneous, but rather depends on the
single space coordinate along the feed gradient.  
A realistic model of the simpler chlorine dioxide-iodine-malonic
acid \cdima (CDIMA) reaction, which is similar to CIMA in terms of its
stationary pattern-forming and dynamical behavior, was put forth
by Lengyel, Rabai, and Epstein (henceforth referred to as LRE) 
\cite{ref:LRE1,ref:LRE2}.
In particular, they demonstrated that the fortuitous choice of
starch as the color indicator for visualizing the patterns in
this system provides the necessary rescaling of the diffusion 
coefficient of the activator (iodide) relative to the inhibitor 
(chlorite) species required for this instability to occur.
Starch is a large molecule that binds to 
the gel matrix of the reactor and is thus effectively immobile.
The reversible binding of iodide with starch results in the purple
starch-triiodide complex and an effective slowing down of the
diffusion rate of iodide.

We begin with the general formulation of an $(N+1)$-component 
reaction-diffusion system with $N$ mobile and a single immobile 
species (as in the CDIMA system), following \cite{ref:PB_92}:
\begin{equation}
  \label{eq:general}
  \frac{\partial \Cvect }{\partial t} 
  = \calDmat \cdot \nabla^2 \Cvect 
         + \Fvect(\Uvect) + \zvect g.
\end{equation}
In the evolution equation above,
\begin{equation}
  \Cvect = \left( \begin{array}{c}
                      \Uvect \\
                      W 
                    \end{array} \right) \, , \qquad
  \calDmat = \left( \begin{array}{cc}
                     \Dmat & {\bf 0} \\
                     {\bf 0}^\top & 0 
                    \end{array} \right)\, , \qquad
  \Fvect = \left( \begin{array}{c}
                      \fvect \\
                      0 
                    \end{array} \right) \, , \qquad
  \zvect = \left( \begin{array}{c}
                      \Rvect \\
                      1 
                    \end{array} \right) \, , 
\end{equation}
$\Uvect$ represents the concentrations of mobile species and $W$ that 
of the immobile species, $\fvect = \fvect(\Uvect; \mbox{\boldmath{$\alpha$}})$ 
describes the reactions among the mobile species with 
control parameters $\mbox{\boldmath{$\alpha$}}$, 
and $g=g(\Uvect, W; \mbox{\boldmath{$\beta$}})$ describes 
the reaction between the mobile species and the immobile species with 
control parameters $\mbox{\boldmath{$\beta$}}$.  
$\Dmat$ is the diffusion matrix for the mobile subsystem,
and $\Rvect$ is a constant stiochiometry vector. We refer the interested reader
to previous works for the description of the LRE model of the CDIMA system.

The steady state of this system in the presence of externally imposed 
boundary feed gradients satisfies:
\begin{eqnarray}
 \fvect(\Uvect_s; \alpha) + \Dmat \cdot 
                                \frac{\partial^2 \Uvect_s}{\partial z^2}
                             & = & 0\, , \\
 g(\Uvect_s, W_s; \beta) & = & 0 
\end{eqnarray}
with boundary conditions $\Cvect_s(0) = \Cvect_0$ and 
$\Cvect_s(\ell) = \Cvect_\ell$.
We will focus on the physically relevant special case 
of scalar diffusion, 
$\Dmat = D \mbox{ \boldmath $I$ } $, 
where $\mbox{ \boldmath $I$ }$ is the identity matrix, as 
the diffusion constants of ions in solution are comparable.
It is clear from the formulation given by Eq.\ \ref{eq:general}, 
that the steady state of the mobile subsystem is independent of the immobile 
reactions.  However, as illustrated in previous 
works \cite{ref:P_92,ref:PB_92,ref:SS_MCC_98,ref:SS_MCC_99}, the 
immobile subsystem affects the stability of the mobile system.
The stability of the steady state to a mode with transverse
wavenumber $k$ ($\hat{k} \perp \hat{z}$) leads to the following linear 
operator:
\begin{equation}
  \lmat_k = \left( \begin{array}{cc}
                         \df & {\bf 0} \\
                         {\bf 0}^\top    & 0 
                      \end{array} \right)  
            - D \left(k^2 + \partial_z^2 \right) \left( \begin{array}{cc}
                         {\bf I} & {\bf 0} \\
                         {\bf 0}^\top & 0
                      \end{array} \right)\,
            + \zvect \, \dg^\top \, ,  
\end{equation}
where $\df = \fvect_{\bf U} $ is the Jacobian derivative of 
$\fvect$ with respect to $\Uvect$, 
and $\dg^\top = \left(g_{\bf U}, g_W \right)$ is the Jacobian 
derivative of $g$ with respect to $(\Uvect, W)$, evaluated at the steady
state.  The eigenvalue 
problem determining the stability of $\Cvect_s(z)$ to 
the mode with wavenumber $k$ is given by 
\begin{equation}
  \lmat_k \cdot \Psivect_k = \lambda_k \Psivect_k
\end{equation}
where $\Psivect = \left(\Psivect^M, \Psi^I \right)$.  
The space $S$ on which the linear operator acts can be written as 
$S = S_M \oplus S_I$, where $S_M$ and $S_I$ are mobile and immobile subspaces, 
respectively. In gradient systems, the basis set spanning $S_M$ is given
by the Cartesian product of $N$ orthogonal
unit vectors $\{\hat{e}_i \}$, $i=1, \ldots, N$,
corresponding to the N mobile species, and $N_z$ basis functions, 
defined on the interval in the $z$-direction, for example given by
$\{ \sin m \pi z/L_z \}$ for Dirichlet boundary conditions, 
where $m=1, \ldots, N_z$.  Similarly, $S_I$ would be
spanned by $\hat{e}_{N+1} \otimes \{ \sin m \pi z/L_z \}$.  We make the 
physical assumption that no instability occurs in the immobile subsystem, 
$\Psivect^M \neq 0$.

A Turing bifurcation occurs when a single $\lambda_k$ goes through zero 
and becomes positive as the parameter $\mbox{\boldmath $\alpha$}$ goes through the 
critical value $\mbox{\boldmath $\alpha_c$}$.
These conditions
\begin{equation}
  \label{eq:conditions}
  \lambda_{k_c} = \left. \frac{d \lambda_k}{d k}\right|_{k = k_c} = 0 
\end{equation}
are satisfied at the onset of the instability, and define the critical 
wavenumber $k_c$.  Note that in the case of scalar diffusion, no such 
instability can occur in the absence of reaction with the immobile species.  
From the above,
\begin{equation}
  \lmat_{k_c} \cdot \Psivect_{k_c} = 0\, , 
\end{equation}
and
\begin{equation}
\label{eq:takeinnerprod}
  \left. \frac{\partial \lmat}{\partial k} \right|_{k_c} 
	\cdot \Psivect_{k_c} 
  +   \lmat_{k_c} \cdot 
	\left. \frac{\partial \Psivect}{\partial k} \right|_{k_c}  = 0\, .
\end{equation}
Taking the inner product of Eq.\ \ref{eq:takeinnerprod}
on the left by $\Psivect^\dagger_{k_c}$, 
the nullvector of $\lmat^\dagger_{k_c}$, we find
\begin{equation}
  {\Psivect_{k_c}^M}^\dagger \cdot \Psivect_{k_c}^M = 0\, ,
\end{equation}
for $k_c \neq 0$. In the absence of the immobile reaction, this 
implies that zero is a defective eigenvalue of 
$\lmat_{k_c}$ at the Turing bifurcation.  
However, it follows that
\begin{equation}
  \Psivect_{k_c}^\dagger \cdot \Psivect_{k_c} = 
  {\Psi_{k_c}^I}^\dagger \Psi_{k_c}^I \neq 0\, 
\end{equation}
since there must exist a nontrivial solution $\Psi^I$ to
\begin{equation}
  g_{\bf U} \cdot \Psivect_{k_c}^M + g_W \Psi^I_{k_c} = 0.
\end{equation}
(Otherwise a nontrivial null vector $\Psivect_{k_c}$ would not exist.)
Hence, zero is not a defective eigenvalue of $\lmat_{k_c}$.  
Furthermore, it can be easily shown that the singularity of 
$\lmat_k$ as a function of parameters $\mbox{\boldmath{$\alpha$}}$ and 
wavenumber $k$ is independent of the immobile reaction \cite{ref:PB_92}.

In this way, the structure of the linear stability of the $(N+1)$-component 
reaction-diffusion system at the Turing bifurcation motivated the 
formulation of a general scheme for appropriately lifting a defective 
eigenvalue problem to a higher dimensional space in which it is no 
longer defective.  Despite this connection, our numerical examples 
of the lifting algorithm in the previous section did not include linear 
stability analysis of the CDIMA system.  This is because the improvement
in accurate calculation of the eigenvector associated with the defective 
eigenvalue is best illustrated when the eigenvalue is exactly known.  
For the gradient CDIMA system, the defective eigenvalue 
$\mu_c \equiv D k_c^2$ is not known apriori.  


\section{Concluding Remarks}
\label{sec:conclude}

The stability analysis of the experimental CDIMA reaction-diffusion system 
has led to the formulation of a new scheme in computational
linear algebra for more accurate numerical solution of defective
eigenvectors.  We have demonstrated improvement in numerical accuracy
by several orders of magnitude using illustrative toy examples.
This scheme should be of utility in physical applications where
accurate eigenvectors of defective matrices are sought, 
as well as of general use in improving accuracy of matrix computations
involving such matrices.

%
%


\end{document}